\pdfoutput=1
\documentclass[aps,prd,amsmath,floats,floatfix, twocolumn,
superscriptaddress,nofootinbib,showpacs,longbibliography]{revtex4-2}

\usepackage[T1]{fontenc}
\usepackage[utf8]{inputenc}
\usepackage{lmodern}
\usepackage{mathrsfs}
\usepackage{amsfonts}

\usepackage{verbatim}

\usepackage[dvipsnames, usenames]{xcolor}
\definecolor{linkcolor}{rgb}{0.0,0.3,0.5}
\usepackage[hypertexnames=false, unicode, colorlinks=true, linkcolor=linkcolor,
citecolor=linkcolor, filecolor=linkcolor,urlcolor=linkcolor,
pdfusetitle]{hyperref}

\usepackage[all]{hypcap}
\usepackage{graphicx}
\usepackage{xspace}
\usepackage{amssymb}

\usepackage[normalem]{ulem} %for \sout
\usepackage{bm} % boldmath

\usepackage{microtype}

\usepackage{blindtext}

\usepackage{orcidlink}

\graphicspath{%
  {figs/}%
}

\DeclareMathAlphabet{\mathpzc}{OT1}{pzc}{m}{it}

\begin{document}

\title{Parameter matching between horizon quasi-local and point-particle definitions at 1PN for quasi-circular and non spinning BBH systems in harmonic gauge}

\newcommand{\Cornell}{\affiliation{Cornell Center for Astrophysics
    and Planetary Science, Cornell University, Ithaca, New York 14853, USA}}
\newcommand\CornellPhys{\affiliation{Department of Physics, Cornell
    University, Ithaca, New York 14853, USA}}
\newcommand\Caltech{\affiliation{Theoretical Astrophysics 350-17, California Institute of
    Technology, 1200 E California Boulevard, Pasadena, CA 91125, USA}}
\newcommand{\AEI}{\affiliation{Max Planck Institute for Gravitational Physics
(Albert Einstein Institute), D-14476 Potsdam, Germany}}
\newcommand{\UMassD}{\affiliation{Department of Mathematics,
    Center for Scientific Computing and Visualization Research,
    University of Massachusetts, Dartmouth, MA 02747, USA}}
\newcommand\UMiss{\affiliation{Department of Physics and Astronomy,
    University of Mississippi, University, MS 38677, USA}}
\newcommand{\Bham}{\affiliation{School of Physics and Astronomy and Institute
    for Gravitational Wave Astronomy, University of Birmingham, Birmingham, B15
    2TT, UK}}
\newcommand{\Perimeter}{\affiliation{Perimeter Institute for Theoretical Physics, Waterloo, ON N2L2Y5, Canada}}

\author{Dongze Sun
\orcidlink{0000-0003-0167-4392}}
\email{dzsun@caltech.edu}
%\thanks{}
\Caltech
\author{Leo C.\ Stein
  \orcidlink{0000-0001-7559-9597}}
\UMiss

% Because hyperref only gets the *last* author, we need to be explicit.
\hypersetup{pdfauthor={Sun et al.}}

\date{\today}

%==========================================================================
\begin{abstract}
We investigate how commonly used parameter definitions in Post-Newtonian (PN) theory compare with those from Numerical Relativity (NR) for binary black hole (BBH) systems. 
In NR, masses and spins of each companion are measured quasi-locally from apparent horizon geometry, whereas in PN they are attributes of point particles defined via asymptotic matching in body zones. 
Although these definitions coincide in the infinite-separation limit, they could differ by finite-separation corrections that matter for precision modeling. 
Working entirely in harmonic gauge, we perform asymptotic matching between each companion's inner zone metric -- obtained from black hole perturbation theory -- and the PN two-body metric, and construct the coordinate transformation that preserves the gauge in the strong field region.
We solve perturbatively for the apparent horizon (AH) on a group of harmonic inertial time slice and compute its quasi-local areal mass from the horizon geometry.
Then we establish the leading order matching between quasi-local (AH based) and PN (point-particle) parameter definitions in harmonic gauge.
We find that on a horizon penetrating harmonic slicing, the AH quasi-local mass agrees with the PN point-particle mass at 1PN order. 
For generic harmonic slicings that deviate from the horizon penetrating condition by a 1PN order perturbation, the AH mass differs from the PN mass also by a 1PN correction.
This parameter matching is crucial for hybridizing PN and NR waveforms and for providing better initial conditions in NR and Cauchy-Characteristic Evolution (CCE) simulations. 
The framework provides a bridge between different descriptions of BBH systems, and it can be extended to spinning and eccentric cases and more general NR gauges.
\end{abstract}

\maketitle

%==========================================================================
\section{Introduction} \label{sec:intro}
Gravitational-wave astronomy is entering a new precision era with the arrival of next-generation observatories.
Ground-based detectors such as Cosmic Explorer (CE) and the Einstein Telescope (ET), together with the space-based Laser Interferometer Space Antenna (LISA), will probe binary black hole coalescences with sensitivities far beyond current capabilities. 
These instruments will record long-duration signals with extremely high signal-to-noise ratios (SNRs) \cite{LISAWaveformWG:2023arg,Barausse:2020mdt}, providing unprecedented access to the dynamics of compact binaries across the inspiral, merger, and ringdown.
As statistical uncertainties on measured source parameters shrink in this high-SNR regime, the burden shifts to theory: waveform models must achieve accuracy levels such that systematic errors remain safely below statistical errors.
This demand places renewed emphasis on the consistency of parameter definitions used across modeling frameworks. 
Even subtle mismatches—such as differences in how masses or spins are defined in post-Newtonian (PN) versus numerical relativity (NR) descriptions—can accumulate over thousands of inspiral cycles and bias astrophysical inference. 
Addressing this challenge requires a parameter-consistent bridge between PN and NR, ensuring that systematic differences in parameter definitions are carefully considered.

For an isolated system, one can define gauge-invariant global quantities such as the Arnowitt–Deser–Misner (ADM) mass and ADM angular momentum, which characterize the energy and angular momentum of the entire spacetime, including contributions from binding energy and gravitational radiation. 
However, in a dynamical binary black hole (BBH) spacetime with gravitational waves, it is generally ill-defined to decompose the ADM mass into “mass of hole 1” and “mass of hole 2” without additional structure.
The difficulty lies in the fact that the self-energy of each black hole is inherently mixed with the system’s binding energy and radiative energy, and any attempt to separate these pieces depends on a choice of gauge. 
NR typically adopts a gauge adapted to the evolution scheme (e.g. moving-puncture or generalized harmonic formulations), whereas post-Newtonian (PN) theory employs the harmonic gauge. 
Thus, while global quantities are unambiguously defined, the assignment of masses and spins to individual black holes in a binary is necessarily gauge-dependent, and PN and NR frameworks make different gauge choices in practice.

Apart from the gauge difference, PN and NR also use different definitions. 
In NR, each black hole’s properties are measured quasi-locally from the geometry of its apparent horizon \cite{Boyle_2019,scheel2025sxscollaborationscatalogbinary}. 
Using techniques from the isolated/dynamical horizon framework \cite{Ashtekar:2003hk,Ashtekar_2004}, one can define the individual mass and angular momentum of a black hole from surface integrals on the AH. 
Since the AH is slicing dependent, the individual mass and angular momentum are also slicing dependent, and they coincide with the ADM quantities only in the infinite-separation limit. 
As the black holes approach each other, each black hole’s horizon mass incorporates effects of the companion’s gravitational field. 
This indicates that the notion of “mass of an individual BH” in a binary is inherently scale-dependent, which agrees with the asymptotic mass at infinite separation, but differs at finite separations due to the presence of binding energy, tidal deformations, and radiations.
The scale-dependent feature of the NR mass was observed in Ref.~\cite{Pook_Kolb_2019}.

In post-Newtonian theory, one treats the black holes as point masses characterized by constant parameters mass and spin, and solves Einstein’s equations as an expansion in $v/c$ or $GM/(c^2r)$ \cite{Blanchet2024}. 
To rigorously connect this idealized point-mass description to an actual black hole, the PN point-mass parameters are defined via an asymptotic matching procedure \cite{poisson2014gravity}: the inner-zone metric near each object is asymptotically expanded in series of $1/r$, and matched in an overlap region (buffer zone) to the global PN metric for the binary system. 
In this way, the constant PN mass parameter is essentially the mass parameter of an isolated BH solution that would produce the same far-field.
At infinite separation, this clearly coincides with each black hole’s physical mass.
However, at finite separation the matching can induce differences from NR, once the two metrics are self-consistently matched. 
Our aim is to quantify this difference at the leading post-Newtonian order.

Comparisons of PN and NR waveforms have indirectly noticed parameter discrepancies. 
Past efforts to hybridize PN and NR waveforms typically assumed that the masses and spins used in PN formulas could be taken directly from the NR horizon masses and spins read from the simulation \cite{Santamaria:2010yb,MacDonald:2011ne,Boyle:2011dy,MacDonald:2012mp,Varma:2018mmi,Sadiq:2020hti,Yoo_2023}. 
This works well to leading order, but as modeling accuracy improved, it became evident that small mismatches in these parameters can visibly affect the waveform phasing. 
Recently, Ref.~\cite{Dongze:2024} demonstrated that allowing slight adjustments (“optimization”) of the PN masses and spins away from the raw NR values yields markedly better PN–NR agreement in hybrid waveforms.
In their study, the PN parameters that minimize the waveform mismatch differ systematically from the NR values. 
This finding suggests that PN and NR are using slightly different conventions for the masses and spins, reinforcing the need for a proper translation in between. 

In this paper, we bridge this gap by deriving an analytic, leading-order parameter mapping between PN and NR definitions of mass for binary black holes.
To do so, we need a spacetime metric that is valid in both inner zone near the horizon ($r\sim r_s \ll b$) and in the near zone where the PN two-body metric is valid ($r_s \ll r \ll \lambda$).
Here $b$ is the separation between the two black holes, $r_s$ is the Schwarzschild radius of the companion black hole that we are interested in, and $\lambda$ is the wavelength of the gravitational radiation.

Refs.\cite{Alvi_2003} and ~\cite{Yunes_2005,Yunes_2006} pioneered an approach to build such global BBH metric, aiming to provide initial data for NR.
Their metric is constructed by matched asymptotic expansions: they asymptotically matched a near zone PN metric to a tidally perturbed BH metric around each black hole in a buffer region where both approximations are valid.
However, in their construction, the coordinates are not enforced to satisfy a single global consistent gauge condition, which can complicate our parameter matching because it's hard to define a time slice in a gauge consistent way.

We work entirely in the harmonic gauge for both the inner zone and near zone to ensure an apple-to-apple comparison without gauge ambiguities. 
We adopt harmonic gauge for simplicity, but this framework can be extended to gauges commonly used in NR, e.g., generalized harmonic formalism used in SpEC.
The inner zone metric in the harmonic gauge is given by Ref.~\cite{Taylor_2008} in the comoving frame with respect to the black hole, which is an isolated BH solution plus tidal multipoles whose leading contributions scale as $\mathcal{O}\left(M^3/b^3\right)$ -- at relative 3PN order. 
Because our goal is a leading-order (1PN) parameter matching, we neglect these higher-order tidal pieces.
The coordinate transformation between the comoving frame and inertial frame is constructed by enforcing the harmonic gauge through the strong field region, the boundary conditions for the gauge equations are provided by asymptotic matching with the PN metric in the buffer region.
Applying this coordinate transformation places the inner zone metric and near zone metric in the same harmonic coordinate system.

We solve perturbatively for the AH on a group of harmonic inertial time slice and compute its quasi-local areal mass from the horizon geometry.
We find that the AH areal mass generally differs from the PN mass at 1PN order under 1PN perturbations, but it reduces to the PN mass when the time slice is horizon penetrating.
For a realistic NR orbital separation (e.g., 10 orbits before merger), if the time slice is not chosen properly near the black hole, the fractional mass difference $\delta m/m$ would be on the order of $10^{-2}$, which is not negligible for the precision requirements of the next generation detectors.

Aligning the PN and NR parameter definitions through this matching has several important implications. 
First, it enables the construction of PN–NR waveform hybrids with consistent physical parameters, and it avoids the risk of overfitting that can bias the parameters.
This should reduce systematic mismatches over long inspirals and give hybrids a more robust physical interpretation. 
Second, our results can inform the setup of NR initial conditions. 
Current NR simulations of BBHs often start from initial data specified by approximate methods (e.g. superposed BHs with PN or effective-one-body estimates for orbital parameters) and involve some iterations to achieve the desired masses and eccentricities after relaxation. 
With a PN–NR parameter mapping, one could choose the initial orbital separation and momenta such that the resulting horizon masses correspond to a given PN binary with parameters $(m_1, m_2)$ at a certain post-Newtonian order. 
This would reduce the initial junk radiation and parameter drift, effectively providing better initial data that already incorporates the PN corrections at a given order. 
Additionally, our framework is formulated in a gauge consistent manner: we work in harmonic coordinates throughout.
In future extensions, the approach can be generalized to other gauge choices by first solving for the coordinate transformation between harmonic coordinates and the desired coordinates, and then locating the AH under the desired slicing condition. 
The framework itself can also be straightforwardly extended to include black-hole spins and orbital eccentricity. 

In this paper, we use lower-case Latin indices from the beginning of the alphabet to denote four-dimensional spacetime quantities, whereas lower-case Latin indices from the middle of the alphabet are spatial.
We work in geometrized units with $G=c=1$ throughout the paper.

The remainder of this paper is organized as follows. 
Sec.~\ref{sec:Def} reviews the definitions of mass in PN theory and in NR. 
Sec.~\ref{sec:matching} presents our asymptotic matching scheme performed in harmonic coordinates.
Sec.~\ref{sec:slicing} specifies the harmonic slicing conditions, and we propose a horizon penetrating harmonic slicing for BBH systems. 
In Sec.~\ref{sec:mass}, we solve for the AH on the matched slice, extract the AH areal mass, and derive the leading-order parameter mapping.
We quantify the magnitude of the correction for typical binary configurations.
Finally, Sec.~\ref{sec:conclusions} concludes the paper with a summary of our key results and an outlook on extensions, including incorporating spins, eccentricities and exploring alternative gauges used in NR.

\section{Mass Definitions in PN and NR}\label{sec:Def}
In this section, we review how mass is defined in the PN and in NR for non-spinning BBH systems, and we clarify how they can be different.

\subsection{PN mass}
In the PN approach, each compact object is idealized as a point particle characterized solely by a constant mass parameter $m_A$ (and spin parameters for spinning case), with $A=1,2$ for a binary.
The gravitational field produced by the point masses is obtained by solving Einstein’s equations expanded in PN orders. 
In harmonic coordinates, where the perturbed metric $h^{ab} = \sqrt{-g}g^{ab} -\eta^{ab}$ (with $\eta_{ab}$ the Minkowski metric) satisfies the harmonic gauge condition $\partial_b h^{ab}=0$. 
The Einstein field equations in harmonic gauge take the form of inhomogeneous flat-space wave equations
\begin{equation}
\Box h^{ab} = 16\pi\tau^{ab}\,
\end{equation}
where
\begin{equation}
\tau^{ab} = |g|T^{ab} + \frac{1}{16\pi}\Lambda^{ab},
\end{equation}
is the effective stress-energy pseudo-tensor of both matter and gravity.
Here $\Lambda^{ab}$ is the nonlinear gravitational source term (quadratic and higher in $h$) arising from self-interactions of the field (see Eq.~(15) of \cite{Blanchet_2002} for full expression), and the stress-energy tensor for the points masses is given by
\begin{equation}
T^{ab}(t,\mathbf{x}) = \sum_{A=1}^2 \frac{m_Au_A^au_A^b}{\sqrt{-g^{cd}(t,\mathbf{y}_A)u_A^c u_A^d}}\frac{\delta^{(3)}[\mathbf{x}-\mathbf{y}_A(t)]}{\sqrt{-g}},
\end{equation}
where $\mathbf{y}_A(t)$ is the trajectory of body $A$ and $u_A^a$ is its 4-velocity.
Then the field solution is given by
\begin{equation}
h^{ab} = 16\pi\Box^{-1}_\text{ret}\tau^{ab},
\end{equation}
where
\begin{equation}
(\Box^{-1}_\text{ret}f)(t,\mathbf{x}) = -\frac{1}{4\pi}\iiint\frac{d^3\mathbf{x'}}{|\mathbf{x}-\mathbf{x'}|}f(t-|\mathbf{x}-\mathbf{x'}|,\mathbf{x'}).
\end{equation}
The PN mass parameters enter the field solutions through this integral.

Because point masses generate singular fields, all these local integrals in the PN iteration require a careful regularization procedure.
Blanchet’s MPM–PN scheme \cite{Blanchet_2014,Blanchet2024} initially employed a Hadamard partie-finie regularization to handle the divergent self-field of each point mass. 
While Hadamard’s method can yield the finite part of most integrals, it was found to break down at the 3PN level, leaving an undetermined constant in the equation of motion.
The resolution of this problem came from using dimensional regularization, which is able to preserve the symmetries of the theory.
In practice, one works in general $d$ spatial dimensions (analytically continued away from $d=3$) so that all integrals are convergent, and defines the physical theory by the limit $d\to 3$ by taking the finite part of any $1/(d-3)$ poles.
It successfully fixed the cutoff ambiguity, yielding a fully self-consistent solution for point-mass binaries at high PN order, and with the mass parameters $m_i$ unambiguously defined.

To rigorously connect this idealized point-mass description to an actual black hole, the PN point-mass parameters are defined alternatively via an asymptotic matching procedure in Poisson's approach \cite{poisson2014gravity}.
Since the PN two-body metric is only valid in the region $r \gg r_\text{s}$, and diverges near the point particle, they replace the singular point-particle with a perturbed black hole solution near the object ($r\ll d$, where $d$ is the separation between two objects) in this object's comoving frame.
Then they solve for the tidal moments of the perturbed black hole solution and the coordinate transformation between the comoving frame and the PN inertial frame by asymptotically expanding the both solutions in comoving frame in series of $1/r$, and matching the coefficients in an overlap region where both the two solutions are valid ($r_\text{s}\ll r\ll d$).
In this way, the constant PN mass parameter is essentially the mass parameter of an perturbed black hole that would produce the same far-field.
At infinite separation or in the limit that the mass of the secondary object tends to zero, the PN mass clearly coincides with the black hole’s Schwarzschild mass.

The horizon absorption of energy should in principle change the mass of each black hole.
PN handles this horizon absorption by absorbing it into the definition of binding energy $E_b$.
In this definition, the PN ADM mass is given by
\begin{equation}
E_\text{ADM} = m_1+m_2+E_b,
\end{equation}
and $E_b$ evolves according to the balance law
\begin{equation}
\frac{d E_b}{dt} = -\left(F_\infty+F_{H,1}+F_{H,2}\right),
\end{equation}
where $F_\infty$ and $F_{H,i}$ are the energy flux to null infinity and the flux into the horizons, respectively.
In this way, the mass parameters entering the PN expressions are kept as constants.

\subsection{NR mass}
In NR, a black hole’s mass is often defined quasi-locally on each time slice using its apparent horizon (AH) -- a marginally outer trapped surface. 
For a non-spinning hole, the quasi-local (irreducible) mass is given by the AH areal mass
\begin{equation}\label{eq:AHmass}
m = \sqrt{\frac{A}{16\pi}},
\end{equation}
where $A$ is the area of the AH of black hole.
This quantity is geometrically invariant on the slice, but it is slicing dependent because the AH itself depends on the chosen time slicing.
By contrast, the event horizon (EH) is a global and gauge invariant null surface, so as the irreducible mass defined from the area of EH. 
However, the EH cannot be determined without knowledge of the entire spacetime evolution, so in practice, the EH based masses is impractical during a simulation.
Consequently, NR uses AH based, quasi-local measurements which depends only on local variables on a given time slice.
The slicing dependence of AH introduces an intrinsic ambiguity, thus the NR quasi-local masses extracted from the AHs need not coincide with the point-particle masses used in PN theory.
Accordingly, any direct comparison between NR and PN requires an explicit, gauge-consistent parameter translation.

\section{Constructing global metric}\label{sec:matching}
In this section we construct the global metric using the matched–asymptotic framework \cite{KOPEIKIN_2004,Taylor_2008}.
We enforce the harmonic gauge $\partial_a(\sqrt{-g}g^{ab})=0$ in the strong field region, so that the global metric and the coordinate transformations are valid and gauge-preserving through the horizon.
We work to leading post-Newtonian order (1PN), and we neglect tidal multipoles whose leading contributions enter at relative 3PN.

\subsection{Region setup and notations}
Consider a binary with component labels $A=1,2$, PN parameters $m_A$, PN trajectories and velocities $\mathbf{y}_A(t)$ and $\mathbf{v}_A(t)$, and separation $b$.
We introduce the usual PN expansion parameter $v=|\mathbf{v_1}-\mathbf{v_2}|\ll1$.
For non-spinning and quasi-circular case, $v^2=(m_1+m_2)/b$ at leading order.
We also have $\partial_tv_1\sim\mathcal{O}(1/b^2)\sim\mathcal{O}(v^4)$, and $\partial_t b\sim\mathcal{O}(v^6)$, so we treat them as constants at 1PN order.
\begin{figure}[thb]
\centering
\includegraphics[width=0.45\textwidth]{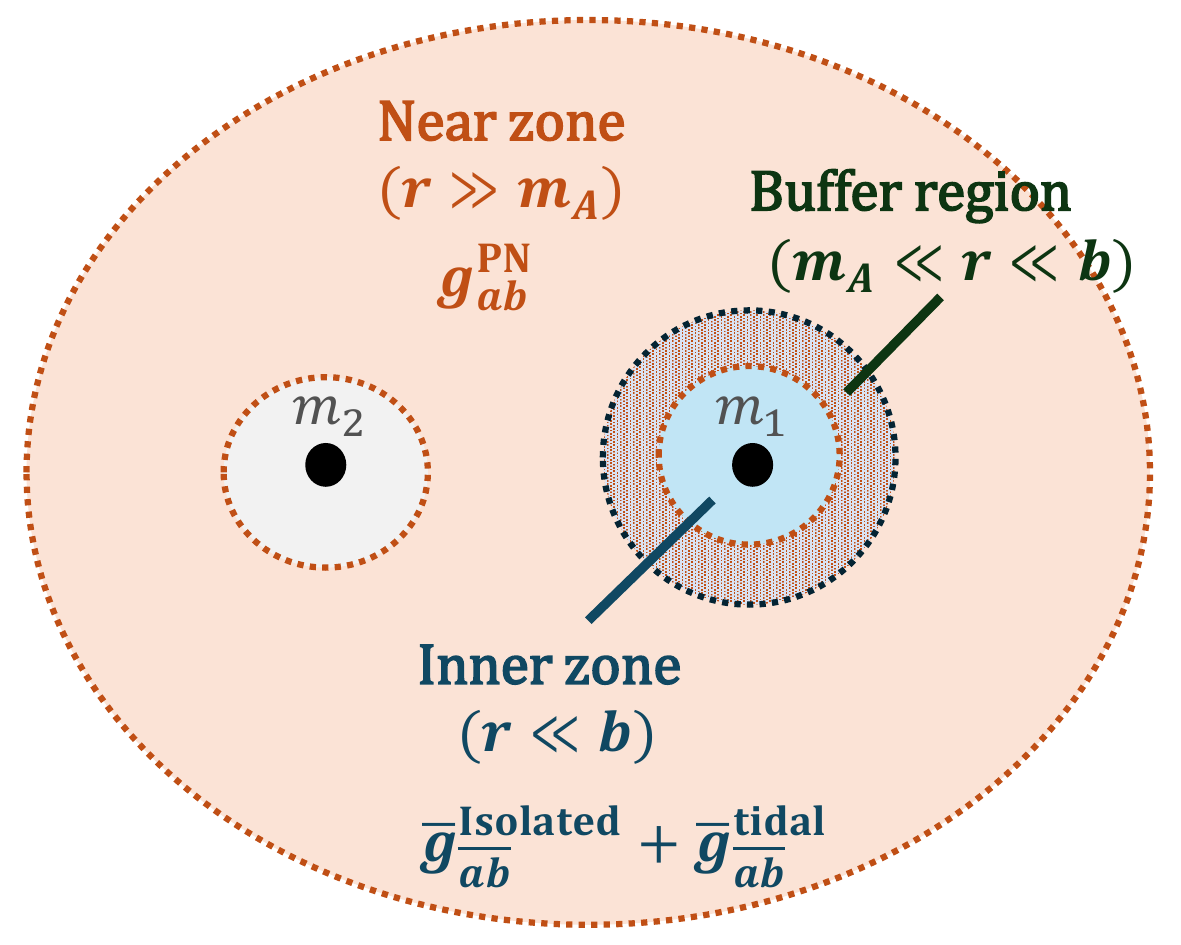}
\caption{A schematic diagram showing the domain setup and symbols.}
\label{fig:schematic}
\end{figure}

Following Ref.~\cite{Yunes_2005}, we divide the global 4-manifold as two inner zones of the black holes ($r_A\ll b$), a near zone ($m_A\ll r_A \ll \lambda$), and a far zone ($r_A\gg\lambda$), as shown in Fig.~\ref{fig:schematic}.
Here $\lambda$ is the wavelength of gravitational radiation, and $r_A = |\mathbf{x}-\mathbf{y_A}|=\sqrt{\delta_{ij}(x^i-y_A^i)(x^j-y_A^j)}\equiv\left((x_A-y_A)(x_A-y_A)\right)$.
Here we use $\left(ab\right)$ to denote the dot product of spatial vectors.
We use $n_A^i=(x^i-y_A^i)/r_A$ to denote the normal vector in $r_A$ direction.
We raise and lower indices of the spatial vector $n_A^i$ and $v_A^i$ with the Euclidean metric $\delta_{ij}$.

We are only interested in the inner zone, where the metric is given as a non-rotating black hole in the harmonic coordinates in the black hole's comoving frame, and the near zone, where the metric is given by the PN two-body metric in the inertial frame.
In order to match the inner zone metric with the near zone metric, we use a buffer region where both descriptions are valid ($m_A\ll r_A\ll b$).

We use the following notation: we use a bar $\bar{}$ to denote coordinates or indices in the comoving frame, and coordinates or indices without the bar are in the inertial frame.
We use superscript or subscript $\text{H}$ for harmonic coordinates, $\text{KS}$ for Kerr-Schild coordinates, and $\text{CS}$ for Cook-Scheel coordinates.

\subsection{Inner zone metric}
In the inner zone of black hole 1, we start from a Kerr-Schild Cartesian coordinate system $\bar{x}_\text{KS}^{\bar{a}} = \left(\bar{t}_\text{KS},\bar{x}_\text{KS}^{\bar{i}}\right)$.
The metric there is taken as an isolated black hole solution plus tidal multipoles \cite{Taylor_2008}. 
The latter start at $\mathcal{O}\left(M^3/b^3\right)$, which are at relative 3PN and are consistently neglected for a leading 1PN mass matching.
Thus in the inner zone we retain only the monopolar field of mass parameter $m^\text{IZ}_1$, which is to be related to the PN parameter $m_1$ by the asymptotic matching.

The inner zone metric of black hole 1 in the Kerr-Schild Cartesian coordinate at 1PN order is
\begin{equation}\label{eq:IZmetric}
\begin{aligned}
g^\text{KS}_{\bar{0}\bar{0}} &= -1+\frac{2m^\text{IZ}_1}{\bar{r}_1^\text{KS}},\\
g^\text{KS}_{\bar{0}\bar{i}} &= \frac{2m^\text{IZ}_1}{\bar{r}_1^\text{KS}}\bar{n}_{1\bar{i}}^\text{KS},\\
g^\text{KS}_{\bar{i}\bar{j}} &=\delta_{\bar{i}\bar{j}}+\frac{2m^\text{IZ}_1}{\bar{r}_1^\text{KS}}\bar{n}_{1\bar{i}}^\text{KS}\bar{n}_{1\bar{j}}^\text{KS}.
\end{aligned}
\end{equation}
Here $\bar{r}_1^\text{KS}=\left(\bar{x}_\text{KS}\bar{x}_\text{KS}\right)$, and $\bar{n}_{1\text{KS}}^{\bar{i}}=\bar{x}_\text{KS}^{\bar{i}}/\bar{r}_1^\text{KS}$.

\subsection{Near zone metric}
In the near zone, the metric can be described by the PN two-body metric in the barycentric harmonic coordinates $x_\text{H}^a = \left(t_\text{H},x_\text{H}^i\right)$ in the inertial frame.
Since our inner zone comoving frame metric is accurate to 2.5PN, we will use the 2.5PN two-body metric to reach the same accuracy in the buffer region.
The 2.5PN metric is given by \cite{Blanchet_1998}.
We substitute $1/r_{2\text{H}}=1/b+\mathcal{O}(v^4)$, keep terms up to $\mathcal{O}(v^2)$, and truncate the metric to $\mathcal{O}(m_1^2/r^2_{1\text{H}})$ order, which is sufficient for our asymptotic matching purpose.
The results reads
\begin{equation}\label{eq:NZmetric}
\begin{aligned}
g^\text{H}_{00}=&-1+\frac{2m_2}{b}+\frac{m_1}{r_{1\text{H}}}\left(2-(n_{1\text{H}}v_1)^2+4v_1^2-\frac{6m_2}{b}\right)\\
&+\frac{m_1^2}{r^2_{1\text{H}}}\left(-2+3(n_{1\text{H}}v_1)^2-v_1^2+\frac{8m_2}{b}\right)\\
&+\mathcal{O}\left(\frac{m_1^3}{r^3_{1\text{H}}},v^3\right),\\
g^\text{H}_{0i}=&-4\frac{m_1}{r_{1\text{H}}}v_{1i}+\frac{m_1^2}{r^2_{1\text{H}}}\left(v_{1i}-(n_{1\text{H}}v_1)n_{1i}^\text{H}\right)\\
&+\mathcal{O}\left(\frac{m_1^3}{r^3_{1\text{H}}},v^3\right),\\
g^\text{H}_{ij}=&\left[1+\frac{2m_2}{b}+\frac{m_1}{r_{1\text{H}}}\left(2-(n_{1\text{H}}v_1)^2+\frac{2m_2}{b}\right)\right.\\
&\left.+\frac{m_1^2}{r^2_{1\text{H}}}\right]\delta_{ij}+\frac{4m_1}{r_{1\text{H}}}v_{1i}v_{1j}+\frac{m_1^2}{r^2_{1\text{H}}}n_{1i}^\text{H}n_{1j}^\text{H}\\
&+\mathcal{O}\left(\frac{m_1^2}{r^2_{1\text{H}}},v^3\right).\\
\end{aligned}
\end{equation}
Note that $g^\text{H}_{ij}$ is not complete at $\mathcal{O}\left(\frac{m_1^2}{r^2_{1\text{H}}}\right)$ order.

\subsection{Coordinate transformation}
Before we match the parameters, it's important to express the inner zone and near zone metrics to the same frame.
Ref.~\cite{Combi_2021} proposed a coordinate transformation between the comoving frame and inertial frame by inserting the PN trajectory of the black hole, but this approach generally does not preserve the desired gauge.
Ref.~\cite{KOPEIKIN_2004} presents a general coordinate transformation by asymptotic matching with the PN metric in the buffer region.
It preserves the harmonic gauge in the weak field region where the metric is flat, but the gauge preserving feature doesn't extend to the strong field region near the horizon.
We construct the coordinate transformation by enforcing the harmonic gauge through the strong field region, and the boundary conditions for the gauge equations are provided by asymptotic matching with the PN metric in the buffer region.

We parameterize the coordinate transformation as
\begin{equation}\label{eq:transformOriginal}
\begin{aligned}
\bar{t}_\text{KS}&=t_\text{H}+\xi^0(t_\text{H},x^i_\text{H}),\\
\left(1-\frac{m_1^\text{IZ}}{\bar{r}_{1\text{KS}}}\right)\bar{x}^{\bar{i}}_\text{KS}&=x^i_\text{H}-y_A^i+\xi^i(t_\text{H},x^i_\text{H}),
\end{aligned}
\end{equation}
here the $\left(1-\frac{m_1^\text{IZ}}{\bar{r}_{1\text{KS}}}\right)$ factor is introduced to bring the comoving Kerr-Schild coordinates to a comoving harmonic coordinates.
We further expand $\xi^0$ as
\begin{equation}\label{eq:xi0}
\xi^0(t_\text{H},x^i_\text{H})=\mathcal{A}(t_\text{H})-\left((x_\text{H}-y_A)v_1\right)+\kappa(t_\text{H},x^i_\text{H}).
\end{equation}
The first two terms are introduced to keep the trajectory, and the third term is introduced to preserve the harmonic gauge.
We have $\dot{\mathcal{A}}\sim\xi^i\sim\mathcal{O}(v^2)$, and $\kappa\sim\mathcal{O}(1)$.

Substituting Eq.~\eqref{eq:xi0} into Eq.~\ref{eq:transformOriginal}, we can solve for the inverse transformation as
\begin{equation}\label{eq:transforminv}
\begin{aligned}
t_\text{H}=&\bar{t}_\text{KS}-\mathcal{A}(\bar{t}_\text{KS})+\left[1-\dot{\mathcal{A}}(\bar{t}_\text{KS})\right]\\
&\cdot\left[\left(1-\frac{m_1^\text{IZ}}{\bar{r}_{1\text{KS}}}\right)\left(\bar{x}_\text{KS}v_1\right)-K(\bar{t}_\text{KS},\bar{x}^{\bar{i}}_\text{KS})\right],\\
x^i_\text{H}=&\left(1-\frac{m_1^\text{IZ}}{\bar{r}_{1\text{KS}}}\right)\bar{x}^{\bar{i}}_\text{KS}+y_A^i\\
&-\xi^i\left(\bar{t}_\text{KS},\left(1-\frac{m_1^\text{IZ}}{\bar{r}_{1\text{KS}}}\right)\bar{x}^{\bar{i}}_\text{KS}\right),
\end{aligned}
\end{equation}
where $K(\bar{t}_\text{KS},\bar{x}^{\bar{i}}_\text{KS})\equiv\kappa\left(t_\text{H}(\bar{t}_\text{KS},\bar{x}^{\bar{i}}_\text{KS}),x^i_\text{H}(\bar{t}_\text{KS},\bar{x}^{\bar{i}}_\text{KS})\right)$.
Eq.~\eqref{eq:transformOriginal} and Eq.~\eqref{eq:transforminv} also gives the transformation of $\bar{r}_\text{1KS}$ and $\bar{n}^{\bar{i}}_\text{1KS}$ as
\begin{equation}\label{eq:transforms3D}
\begin{aligned}
\bar{r}_\text{1KS}=&r_{1\text{H}}+(n_{1\text{H}}\xi)+m_1,\\
\bar{n}^{\bar{i}}_\text{1KS}=&\left[1-\frac{(n_{1\text{H}}\xi)}{r_{1\text{H}}}\right]n^i_\text{1H}+\frac{\xi^i}{r_{1\text{H}}}.
\end{aligned}
\end{equation}

The harmonic condition reads
\begin{equation}\label{eq:harmonic}
g_\text{KS}^{\bar{a}\bar{b}}\left(\partial_{\bar{a}}\partial_{\bar{b}}x^c_\text{H}-\Gamma^{\bar{d}}_{\text{KS}\bar{a}\bar{b}}\partial_{\bar{d}}x^c_\text{H}\right)=0.
\end{equation}
Substituting Eq.~\eqref{eq:transforminv}, we get
\begin{equation}\label{eq:harmonicK}
\begin{aligned}
\bar{\nabla}^2K&-\frac{2m_1^\text{IZ}}{\bar{r}_{1\text{KS}}}\bar{n}_{1\text{KS}}^{\bar{i}}\bar{n}_{1\text{KS}}^{\bar{j}}\partial_{\bar{i}}\partial_{\bar{j}}K\\
&-\frac{2m_1^\text{IZ}}{\bar{r}^2_{1\text{KS}}}\left(1+\bar{n}_{1\text{KS}}^{\bar{i}}\partial_{\bar{i}}K\right)=\mathcal{O}(v^3),
\end{aligned}
\end{equation}
\begin{equation}\label{eq:harmonicxi}
\left(1-\frac{m_1^\text{IZ}}{\bar{r}_{1\text{KS}}}\right)\bar{\nabla}^2\xi^k-\frac{m_1^\text{IZ}}{\bar{r}_{1\text{KS}}}\bar{n}_{1\text{KS}}^{\bar{i}}\bar{n}_{1\text{KS}}^{\bar{j}}\partial_{\bar{i}}\partial_{\bar{j}}\xi^k=\mathcal{O}(v^3),
\end{equation}
where $\nabla^2$ is the flat Laplacian.
We can further decompose $\xi^i$ as
\begin{equation}
\xi^i(t_\text{H},x^j_\text{H})=B(t_\text{H},x^j_\text{H})n_{1\text{H}}^i+D(t_\text{H},x^j_\text{H})v_1^i,
\end{equation}
then we substitute it into Eq.~\eqref{eq:harmonic} to get
\begin{equation}\label{eq:harmonicB}
\bar{\nabla}^2B-\frac{m_1^\text{IZ}}{\bar{r}_{1\text{KS}}}\bar{n}_{1\text{KS}}^{\bar{i}}\bar{n}_{1\text{KS}}^{\bar{j}}\partial_{\bar{i}}\partial_{\bar{j}}B-\frac{2B}{\bar{r}_{1\text{KS}}^2}=\mathcal{O}(v^3),
\end{equation}
\begin{widetext}
\begin{equation}\label{eq:harmonicD}
\begin{aligned}
&\frac{2B}{\bar{r}_{1\text{KS}}^2}\left(v_1\bar{n}_{1\text{KS}}\right)-\frac{2v_1^{\bar{i}}\partial_{\bar{i}}B}{\bar{r}_{1\text{KS}}}+\frac{m_1^\text{IZ}}{\bar{r}_{1\text{KS}}}\left(v_1\bar{n}_{1\text{KS}}\right)\bar{n}_{1\text{KS}}^{\bar{i}}\bar{n}_{1\text{KS}}^{\bar{j}}\partial_{\bar{i}}\partial_{\bar{j}}B+\bar{n}_{1\text{KS}}^{\bar{i}}v_1^{\bar{j}}\partial_{\bar{i}}\partial_{\bar{j}}B+\frac{2\left(v_1\bar{n}_{1\text{KS}}\right)\bar{n}_{1\text{KS}}^{\bar{i}}\partial_{\bar{i}}B}{\bar{r}_{1\text{KS}}}-v_1^{\bar{i}}v_1^{\bar{j}}\partial_{\bar{i}}\partial_{\bar{j}}D\\
&+\frac{m_1^\text{IZ}v_1^2}{\bar{r}_{1\text{KS}}}\bar{n}_{1\text{KS}}^{\bar{i}}\bar{n}_{1\text{KS}}^{\bar{j}}\partial_{\bar{i}}\partial_{\bar{j}}D=\mathcal{O}(v^4).
\end{aligned}
\end{equation}
\end{widetext}

The harmonic condition Eq.~\eqref{eq:harmonic} does not constrain $\mathcal{A}$ at 1PN order, so we will need to get information of $\mathcal{A}$ and acquire boundary conditions for $K, B$ and $D$ by asymptotically matching the inner zone metric with the PN near zone metric in the buffer region.

\subsection{Asymptotic matching}
In order to asymptotically match the inner zone metric with the PN near zone metric, we first need to transform the inner zone metric to the inertial frame in the harmonic coordinates
\begin{equation}\label{eq:transformedIZmetric}
g^\text{H}_{ab}(t_\text{H},x^j_\text{H})=g^\text{KS}_{\bar{c}\bar{d}}\frac{\partial \bar{x}^{\bar{c}}_\text{KS}}{\partial x^a_\text{H}}\frac{\partial \bar{x}^{\bar{d}}_\text{KS}}{\partial x^b_\text{H}}.
\end{equation}
The Jacobian is given by Eq.~\eqref{eq:transformOriginal} and \eqref{eq:transforminv} as
\begin{equation}\label{eq:Jacobian}
\begin{aligned}
\partial^{\bar{0}}_0=&1+\dot{\mathcal{A}}+v_1^2+\partial_t\kappa-v_1^i\partial_i \kappa,\\
\partial^{\bar{0}}_i=&-v_{1i}+\partial_i\kappa,\\
\partial^{\bar{i}}_0=&-\left(1+\frac{m_1^\text{IZ}}{r_{1\text{H}}}\right)v_1^i+\frac{m_1^\text{IZ}}{r_{1\text{H}}}(v_1n_{1\text{H}})n_{1\text{H}}^i,\\
\partial^{\bar{i}}_j=&\left[1+\frac{B}{r_{1\text{H}}}+\frac{m_1^\text{IZ}}{r_{1\text{H}}}\left(1-\frac{(v_1n_{1\text{H}})D}{r_{1\text{H}}}\right)\right]\delta^i_j\\
&+n^i_{1\text{H}}\partial_jB-\frac{m_1^\text{IZ}(v_1n_{1\text{H}})}{r_{1\text{H}}}n^i_{1\text{H}}\partial_jD\\
&+\left(1+\frac{m_1^\text{IZ}}{r_{1\text{H}}}\right)v^i_1\partial_jD+\left[-\frac{B}{r_{1\text{H}}}\right.\\
&\left.+\frac{2m_1^\text{IZ}(v_1n_{1\text{H}})D}{r^2_{1\text{H}}}-\frac{m_1^\text{IZ}}{r_{1\text{H}}}\left(1-\frac{(v_1n_{1\text{H}})D}{r_{1\text{H}}}\right)\right]n^i_{1\text{H}}n^\text{H}_{1j}\\
&-\frac{m_1^\text{IZ}D}{r^2_{1\text{H}}}\left(n^i_{1\text{H}}v_{1j}+v^i_1n^\text{H}_{1j}\right).
\end{aligned}
\end{equation}
We apply the Jacobian and Eq.~\eqref{eq:transforms3D} to the inner zone metric Eq.~\eqref{eq:IZmetric}, and expand the results in series of $1/r_{1\text{H}}$, and keep the terms up to the second order.
The full transformed metric is very lengthy, so we will just give the useful parts.
The $\delta_{ij}$ component of the transformed metric reads
\begin{equation}
\begin{aligned}
g^{\text{H}(\delta)}_{ij}=&1+\frac{2m^\text{IZ}_1+2B}{r_{1\text{H}}}\\
&+\frac{m^{2}_{1\text{IZ}}+2m_1^\text{IZ}B+B^2-2m_1^\text{IZ}D(n_{1\text{H}}v_1)}{r^2_{1\text{H}}}\\
&+\mathcal{O}\left(\frac{m_1^3}{r^3_{1\text{H}}},v^3\right).\\
\end{aligned}
\end{equation}
By comparing with Eq.~\eqref{eq:NZmetric}, we immediately get
\begin{equation}\label{eq:BCBD}
m^\text{IZ}_1=m_1,\quad B\doteq\frac{m_2}{b}r_{1\text{H}},\quad D\doteq\frac{1}{2}r_{1\text{H}}(v_1n_{1\text{H}}),
\end{equation}
where $\doteq$ denotes equality in the buffer region (boundary condition).
For $B$, the boundary condition is spherically symmetric, so we look for a spherical symmetric solution to Eq.~\eqref{eq:harmonicB}.
The general spherical symmetric solution is 
\begin{equation}
B=\beta_1 r_{1\text{H}}+\beta_2\left(-1-\frac{r_{1\text{H}}}{r_{1\text{H}}-m1}+\frac{2r_{1\text{H}}}{m_1}\ln\frac{r_{1\text{H}}}{r_{1\text{H}}-m_1}\right),
\end{equation}
where the second term scales like $\mathcal{O}(m_1^2/r_{1\text{H}}^2)$, so it can not be constrained by the current 2.5PN metric whose spatial components are not complete at $\mathcal{O}(m_1^2/r_{1\text{H}}^2)$ order.
So we will choose $\beta_1=m2/d$, and $\beta_2=0$.
This gives
\begin{equation}\label{eq:B}
B=\frac{m_2}{b}r_{1\text{H}}.
\end{equation}
Eq.~\eqref{eq:harmonicD} can then be simplified by substituting $B$, which gives
\begin{equation}
\begin{aligned}
\nabla^2D-\frac{m_1^\text{IZ}}{\bar{r}_{1\text{KS}}}\bar{n}_{1\text{KS}}^{\bar{i}}\bar{n}_{1\text{KS}}^{\bar{j}}\partial_{\bar{i}}\partial_{\bar{j}}D=\mathcal{O}(v^3).
\end{aligned}
\end{equation}
Based on the boundary condition Eq.~\eqref{eq:BCBD}, we look for a $l=1$ solution.
The general $l=1$ solution can be written as
\begin{equation}
D=(\delta_1+\delta_2 r_{1\text{H}})(n_{1\text{H}}u),
\end{equation}
where $\mathbf{u}$ is a constant vector.
The boundary condition Eq.~\eqref{eq:BCBD} gives $\delta_1=0,\delta_2=\frac{1}{2}$, and $\mathbf{u}=\mathbf{v_1}$.
So we have
\begin{equation}\label{eq:D}
D=\frac{1}{2}r_{1\text{H}}(v_1n_{1\text{H}}).
\end{equation}

To simplify the matching for $\kappa$ and $\mathcal{A}$, we first study the gauge constraint Eq.~\eqref{eq:harmonicK}.
This equation is independent of $\bar{t}_\text{KS}$, so we can decompose $K$ as $K(\bar{t}_\text{KS},\bar{x}^{\bar{i}}_\text{KS})=K_1(\bar{t}_\text{KS})K_2(\bar{x}^{\bar{i}}_\text{KS})$.
Given that both the inner zone metric and near zone metric have no explicit dependence on time, and that $K_1(\bar{t}_\text{KS})$ enters directly in the Jacobian Eq.~\eqref{eq:Jacobian}, then $K_1(\bar{t}_\text{KS})$ must be constant.
So $K=K(\bar{x}^{\bar{i}}_\text{KS})$.
For simplicity, we first look for a spherical symmetric solution to Eq.~\eqref{eq:harmonicK}.
The general solution is
\begin{equation}\label{eq:Kansatz}
K=2m_1\ln\frac{\bar{r}_{1\text{KS}}-2m_1}{k_1}+k_2\ln\frac{\bar{r}_{1\text{KS}}-2m_1}{\bar{r}_{1\text{KS}}},
\end{equation}
where the $k_1$ coefficient is just an arbitrary constant time shift, so we can choose it to be $2m_1$ to make the unit correct.
We substitute the coordinate transformation Eq.~\eqref{eq:transforms3D} into ansatz Eq.~\eqref{eq:Kansatz} to get the ansatz for $\kappa$
\begin{equation}\label{eq:kappaansatz}
\begin{aligned}
\kappa=&2m_1\ln\frac{r_{1\text{H}}\left(1+\frac{m_2}{b}+\frac{1}{2}(n_{1\text{H}}v_1)\right)-m_1}{2m_1}\\
&+k_2\ln\frac{r_{1\text{H}}-2m_1\left(1-\frac{m_2}{b}-\frac{1}{2}(n_{1\text{H}}v_1)\right)}{r_{1\text{H}}},
\end{aligned}
\end{equation}
and then substitute the $\kappa$ along with \eqref{eq:B} and \eqref{eq:D} into the transformed metric Eq.~\eqref{eq:transformedIZmetric}.
The $\mathcal{O}(1)$ coefficient of the tt component is
\begin{equation}
g^{\text{H}(\mathcal{O}(1))}_{ij}=-1-2\dot{\mathcal{A}}-v_1^2.
\end{equation}
Comparing it with the tt component of Eq.~\eqref{eq:NZmetric}, we have
\begin{equation}\label{eq:A}
\dot{\mathcal{A}}=-\frac{1}{2}v_1^2-\frac{m_2}{b}.
\end{equation}
Then the $\mathcal{O}(m_1^2/r^2_{1\text{H}})$ coefficient of the tt component reduces to
\begin{equation}
\begin{aligned}
g^{\text{H}(\mathcal{O}(m_1^2/r^2_{1\text{H}}))}_{ij}=&\frac{m_1^2}{r^2_{1\text{H}}}\left(-2+2k_2(n_{1\text{H}}v_1)\right.\\
&\left.+3(n_{1\text{H}}v_1)^2-v_1^2+\frac{8m_2}{b}\right).
\end{aligned}
\end{equation}
Comparing with the Eq.~\eqref{eq:NZmetric}, we have $k_2=0$.
So
\begin{equation}\label{eq:kappa}
\kappa=2m_1\ln\frac{r_{1\text{H}}\left[1+\frac{m_2}{b}+\frac{1}{2}(n_{1\text{H}}v_1)\right]-m_1}{2m_1}.
\end{equation}

We use $B, D, \dot{\mathcal{A}}$ and $\kappa$ from Eq.~\eqref{eq:B}, \eqref{eq:D}, \eqref{eq:A} and \eqref{eq:kappa} to transform the inner zone metric Eq.~\eqref{eq:IZmetric}, the results agrees exactly with the near zone metric Eq.~\eqref{eq:NZmetric} for the available orders in $m_1/r_{1\text{H}}$ up to $\mathcal{O}(v^2)$.
This suggests that the $l>0$ modes of $K$ vanishes at 1PN order, which agrees our spherical symmetric ansatz Eq.~\eqref{eq:Kansatz}.

In conclusion, the harmonic gauge-preserving coordinate transformation from the PN inertial frame to the comoving Kerr-Schild coordinate is given by
\begin{equation}\label{eq:transformFinal}
\begin{aligned}
\bar{t}_\text{KS}=&\left(1-\frac{1}{2}v^2_1-\frac{m_2}{b}\right)t_\text{H}-\left((x_\text{H}-y_A)v_1\right)\\
&+2m_1\ln\frac{r_{1\text{H}}\left[1+\frac{m_2}{b}+\frac{1}{2}(n_{1\text{H}}v_1)\right]-m_1}{2m_1},\\
\bar{x}^{\bar{i}}_\text{KS}=&\frac{r_{1\text{H}}}{r_{1\text{H}}-m_1}\left[\left(1+\frac{m_2}{b}\right)\left(x^i_\text{H}-y_A^i\right)\right.\\
&\left.+\frac{1}{2}r_{1\text{H}}(n_{1\text{H}v_1})v_1^i\right],\\
\bar{r}_{1\text{KS}}=&r_{1\text{H}}\left(1+\frac{1}{2}(n_{1\text{H}}v_1)^2+\frac{m_2}{b}\right)+m_1,\\
\bar{n}^{\bar{i}}_\text{1KS}=&\left(1-\frac{1}{2}(n_{1\text{H}}v_1)^2\right)n_{1\text{H}}^i+\frac{1}{2}(n_{1\text{H}}v_1)v^i_1.
\end{aligned}
\end{equation}
The inverse transformation is
\begin{equation}\label{eq:transformFinalinv}
\begin{aligned}
t_\text{H}=&\left(1+\frac{1}{2}v^2_1+\frac{m_2}{b}\right)\left[\bar{t}_\text{KS}+\left(\bar{r}_{1\text{KS}}-m_1\right)\left(\bar{n}_\text{KS}v_1\right)\right.\\
&\left.-2m_1\ln\frac{\bar{r}_{1\text{KS}}-2m_1}{2m_1}\right],\\
x^i_\text{H}=&\left(\bar{r}_{1\text{KS}}-m_1\right)\left[\left(1-\frac{m_2}{b}\right)\bar{n}^{\bar{i}}_\text{1KS}-\frac{1}{2}\left(\bar{n}_\text{KS}v_1\right)v_1^i\right]\\
&+y_A^i,\\
r_{1\text{H}}=&\left(1-\frac{1}{2}(\bar{n}_\text{1KS}v_1)^2-\frac{m_2}{b}\right)\left(\bar{r}_{1\text{KS}}-m_1\right),\\
n_{1\text{H}}^i=&\left(1+\frac{1}{2}(\bar{n}_\text{1KS}v_1)^2\right)\bar{n}^{\bar{i}}_\text{1KS}-\frac{1}{2}(\bar{n}_\text{KS}v_1)v^i_1.
\end{aligned}
\end{equation}

\section{Slicing condition}\label{sec:slicing}
The apparent horizon is slicing dependent, but the slicing conditions that satisfy the harmonic condition is not unique.
We can add any function $F(\bar{t}_\text{KS},\bar{x}^{\bar{i}}_\text{KS})$ to the harmonic time $t_\text{H}$ in Eq.~\eqref{eq:transformFinalinv} as long as $F$ satisfies the harmonic condition Eq.~\eqref{eq:harmonic}.
For simplicity, we focus on the static case, so $F=F(\bar{x}^{\bar{i}}_{1\text{KS}})$.
Putting it into Eq.~\eqref{eq:harmonic}, we get
\begin{equation}
\bar{\nabla}^2F-\frac{2m_1}{\bar{r}_{1\text{KS}}}\bar{n}_{1\text{KS}}^{\bar{i}}\bar{n}_{1\text{KS}}^{\bar{j}}\partial_{\bar{i}}\partial_{\bar{j}}F-\frac{2m_1}{\bar{r}_{1\text{KS}}^2}\bar{n}_{1\text{KS}}^{\bar{i}}\partial_{\bar{i}}F=0.
\end{equation}
The general spherically symmetric solution is
\begin{equation}
F=f_1\ln\frac{\bar{r}_{1\text{KS}}-2m_1}{\bar{r}_{1\text{KS}}}+f_2,
\end{equation}
where $f_2$ is just an arbitrary constant time shift that we are setting to zero.

We note that when
\begin{equation}
f_1=2m_1\left(1+\frac{1}{2}v^2_1+\frac{m_2}{b}\right),
\end{equation}
the harmonic time slice becomes
\begin{equation}\label{eq:tCS}
\begin{aligned}
t_\text{CS}=&\left(1+\frac{1}{2}v^2_1+\frac{m_2}{b}\right)\left[\bar{t}_\text{KS}+\left(\bar{r}_{1\text{KS}}-m_1\right)\left(\bar{n}_\text{KS}v_1\right)\right.\\
&\left.+2m_1\ln\frac{2m_1}{\bar{r}_{1\text{KS}}}\right]=const.,
\end{aligned}
\end{equation}
which is regular through the horizon.
This is the generalized case of the Cook-Scheel slicing \cite{Cook_1997}, which is the unique harmonic and horizon penetrating slicing for an isolated single black hole.

To study the dependence of the AH quasi-local mass on the slicing condition, we perturb the generalized Cook-Scheel slicing condition in Eq.~\eqref{eq:tCS} by setting $f_1=2m_1\left(1+\frac{1}{2}v^2_1+\frac{m_2}{b}\right)-2m_1\delta v^2$.
The slicing condition then becomes
\begin{equation}\label{eq:slicing}
\begin{aligned}
T=&\left(1+\frac{1}{2}v^2_1+\frac{m_2}{b}\right)\left[\bar{t}_\text{KS}+\left(\bar{r}_{1\text{KS}}-m_1\right)\left(\bar{n}_\text{KS}v_1\right)\right.\\
&\left.+2m_1\ln\frac{2m_1}{\bar{r}_{1\text{KS}}}\right]-2m_1\delta v^2\ln\frac{\bar{r}_{1\text{KS}}-2m_1}{\bar{r}_{1\text{KS}}}=const..
\end{aligned}
\end{equation}

We will solve perturbatively for the AH on this slicing condition and compute its quasi-local areal mass from the horizon geometry in the next section.

\section{AH quasi-local mass}\label{sec:mass}
The apparent horizon of a black hole is defined as an outer marginally trapped surface on a given time slice.
It is a closed 2-surface whose outgoing null normal has zero expansion.
If one combines the AH on each time slice into a 3-dimensional surface, this worldtube will depend on the slicing.
As we will see, this slicing dependent feature of the AH will bring ambiguity into the quasi-local mass defined from the AH geometry.

\subsection{Finding the apparent horizon}
Here we solve for the AH on the given time slice $T=const.$ in Eq.~\eqref{eq:slicing}.
The future-pointing unit timelike normal to this time slice is given by
\begin{equation}
\tau_a=-\alpha D_aT,
\end{equation}
where $D_a$ is the covariant derivative with respect to the 4-metric $g_{ab}$, and
\begin{equation}
\alpha=(-g^{ab}D_aTD_bT)^{-1/2}.
\end{equation}
The induced 3-metric on this time slice is given by
\begin{equation}
g^{(3)}_{ab}=g_{ab}+\tau_a\tau_b.
\end{equation}
Let $s^a$ be the unit outward pointing normal to the AH 2-surface in this time slice, so that
\begin{equation}\label{eq:basic}
s^a\tau_a=0, \quad s^as_a=1,\quad \tau^a\tau_a=-1,
\end{equation}
then the 2-metric on the AH 2-surface is
\begin{equation}\label{eq:2metric}
g^{(2)}_{ab}=g_{ab}+\tau_a\tau_b-s_as_b,
\end{equation}
and
\begin{equation}
k^a=\frac{1}{\sqrt{2}}(\tau^a+s^a)
\end{equation}
is the future-pointing null geodesic congruence whose projection on the 3D time slice is orthogonal to the AH 2-surface.
Note that $k^a$ is only defined on the AH 2-surface, its extension away from the 2-surface is not defined.
The AH is defined by the vanishing of expansion on the 2-surface, which is
\begin{equation}\label{eq:AHOriginal}
\Theta\equiv g^{(2)ab}D_ak_b=0.
\end{equation}
Since the covariant derivative of $k_a$ is projected onto the 2-surface by $g^{(2)ab}$, no knowledge of the extension of $k^a$ away from the 2-surface is required.
Only if $k^a$ is affinely parametrized, i.e., $k^aD_ak^b=0$ in the neighborhood of the 2-surface, Eq.~\eqref{eq:AHOriginal} reduces to $D_ak^a=0$.

Note that $\tau^aD_b\tau_a=0$ and $s^aD_bs_a=0$ from Eq.~\eqref{eq:basic}, Eq.~\eqref{eq:AHOriginal} can be simplified as
\begin{equation}\label{eq:AH}
\Theta= \frac{1}{\sqrt{2}}\left[g^{ab}D_a(\tau_b+s_b)+\tau^a\tau^bD_as_b-s^as^bD_a\tau_b\right]=0.
\end{equation}
We will use this equation for the remaining calculations.

In order to get $s^a$, we parametrize the AH 2-surface in the comoving Kerr-Schild coordinates by
\begin{equation}\label{eq:R}
\begin{aligned}
\mathcal{R}=& \bar{r}_{1\text{KS}}-2m_1-R(\bar{\theta},\bar{\phi})\\
=&\bar{r}_{1\text{KS}}-2m_1-\sum_{\bar{l},\bar{m}}\bar{R}_{\bar{l}\bar{m}}Y^{\bar{l}\bar{m}}(\bar{\theta},\bar{\phi})=0,
\end{aligned}
\end{equation}
with $(\bar{\theta},\bar{\phi})$ defined by 
\begin{equation}
\bar{x}^{\bar{i}}_{\text{KS}}=\bar{r}_{1\text{KS}}\left(\sin(\bar{\theta})\cos(\bar{\phi}),\sin(\bar{\theta})\sin(\bar{\phi}),\cos(\bar{\phi})\right).
\end{equation}
Here $\bar{R}_{\bar{l}\bar{m}}\sim\mathcal{O}(v^2)$, given that we are on a time slice that is perturbed from the horizon-penetrating one by $\mathcal{O}(v^2)$.
Then we have
\begin{equation}
s^a=\frac{g^{(3)ab}D_b\mathcal{R}}{\sqrt{g^{(3)cd}D_c\mathcal{R}D_d\mathcal{R}}}.
\end{equation}

Eq.~\eqref{eq:AH} is a scalar equation which does not transform under coordinate transformations, so we solve it in the comoving Kerr-Schild coordinates, which simplifies the calculation significantly compared with working in the inertial harmonic coordinates.
We substitute the slicing Eq.~\eqref{eq:slicing} and the comoving metric Eq.~\eqref{eq:IZmetric} into Eq.~\eqref{eq:AH}, project it onto $\bar{l},\bar{m}$ modes, and truncate terms to $\mathcal{O}(v^2)$ and $\mathcal{O}(R)$, we obtain
\begin{equation}
\begin{aligned}
\bar{R}_{\bar{0}\bar{0}}=& -\sqrt{\pi}m_1\delta v^2+\mathcal{O}(v^3),\\
\bar{R}_{\bar{l}>0,\bar{m}}=&\mathcal{O}(v^3),
\end{aligned}
\end{equation}
so the AH is given by
\begin{equation}\label{eq:AHsolution}
\bar{r}_{1\text{KS}}=2m_1-\frac{m_1}{2}\delta v^2.
\end{equation}
For the horizon penetrating slicing, i.e., $\delta=0$, the AH reduces to the Kerr-Schild horizon at $\bar{r}_{1\text{KS}}=2m_1$.

\subsection{Quasi-local mass}
The AH quasi-local mass is defined from the area of the AH by Eq.~\eqref{eq:AHmass}.
The AH area $A$ is given by
\begin{equation}
A=\int_0^{2\pi}d\bar{\phi}\int_0^{\pi}d\bar{\theta}\sqrt{g^{(2)}_{\bar{\theta}\bar{\theta}}g^{(2)}_{\bar{\phi}\bar{\phi}}-g^{(2)2}_{\bar{\theta}\bar{\phi}}},
\end{equation}
where
\begin{equation}
g^{(2)}_{\bar{\theta i}\bar{\theta j}}=g^{(2)}_{\bar{a}\bar{b}}\partial_{\bar{\theta i}}\bar{x}^{\bar{a}}_{1\text{KS}}\partial_{\bar{\theta j}}\bar{x}^{\bar{b}}_{1\text{KS}}.
\end{equation}
Note that from Eq.~\eqref{eq:slicing}, we have $\bar{t}_\text{KS}=const.-\left(\bar{r}_{1\text{KS}}-m_1\right)\left(\bar{n}_\text{KS}v_1\right)+...$, so the Jacobian is
\begin{equation}
\begin{aligned}
\partial_{\bar{\theta}}\bar{x}^{\bar{a}}=&\bar{r}_{1\text{KS}}\left((1-\frac{m_1}{\bar{r}_{1\text{KS}}})v_1\cos(\bar{\theta})\sin(\bar{\phi}),\cos(\bar{\theta})\cos(\bar{\phi}),\right.\\
&\left.\cos(\bar{\theta})\sin(\bar{\phi}),-\sin(\bar{\theta})\right),\\
\partial_{\bar{\phi}}\bar{x}^{\bar{a}}=&\bar{r}_{1\text{KS}}\left((1-\frac{m_1}{\bar{r}_{1\text{KS}}})v_1\sin(\bar{\theta})\cos(\bar{\phi}),-\sin(\bar{\theta})\sin(\bar{\phi}),\right.\\
&\left.\sin(\bar{\theta})\cos(\bar{\phi}),0\right).
\end{aligned}
\end{equation}
Substituting the AH \eqref{eq:AHsolution} and the 2-metric \eqref{eq:2metric}, we get the AH quasi-local mass from Eq.~\eqref{eq:AHmass} as
\begin{equation}
m_1^\text{AH}=m_1\left(1-\frac{1}{4}\delta v^2\right).
\end{equation}

For the horizon penetrating slicing, i.e., $\delta=0$, the AH quasi-local mass agrees with the PN mass at 1PN order.
But for a general harmonic slicing condition that is perturbed from the horizon penetrating one by 1PN order at $\mathcal{O}(v^2)$, the AH quasi-local mass is also perturbed from the PN mass by $\mathcal{O}(v^2)$.
For a realistic NR orbital separation (e.g., 10 orbits before merger), $\mathcal{O}(v^2)$ would be on the order of $10^{-2}$, which is not negligible for the precision requirements of the next generation detectors.

\section{Conclusions}\label{sec:conclusions}
We have provided a gauge-consistent bridge between apparent horizon quasi-local and PN point-particle parameter definitions for non-spinning, quasi-circular BBH systems at leading PN order, working entirely in harmonic gauge. 
By constructing a harmonic and gauge-preserving map between the inner zone black hole metric and the near zone PN two-body metric, we constructed a global harmonic metric for BBH, and we generalized the Cook-Scheel harmonic horizon penetrating slicing to the BBH case.
We located the apparent horizon on harmonic time slices, and derived the leading order relation between the AH areal mass and the PN mass parameter at 1PN order. 
We find that on a horizon penetrating harmonic slicing (the generalized Cook–Scheel slice), the AH quasi-local mass agrees with the PN point-particle mass at 1PN order. 
For generic harmonic slicings that deviate from the horizon penetrating condition by an $\mathcal{O}(v^2)$ perturbation, the AH mass differs from the PN mass by a fractional amount $\frac{\Delta m_1}{m_1}=-\frac{1}{4}\delta v^2$.
For typical orbital velocities at realistic NR separations (e.g., 10 orbits before merger), this amounts to $\mathcal{O}(10^{-2})$ corrections, which is not negligible at the precision demanded by next generation detectors.

Our work neglected tidal multipoles entering at relative 3PN, and restricted to non-spinning, quasi-circular binaries. 
Some natural extensions would include: (i) incorporating spin and eccentricity to obtain mappings for both mass and angular momentum; (ii) adding leading tidal effects and extend to higher PN orders; (iii) translating the mapping from harmonic to gauges more commonly used in NR (e.g., damped harmonic in SpEC) via explicit coordinate transforms that preserves the new gauge.

Extending the matching procedure to the actual NR gauges would enable PN–NR waveform hybridizations to be performed with parameter consistent inputs, reducing the need for ad-hoc optimization of PN masses to fit NR and thereby decreasing the risk of parameter bias over long inspirals. 
It also suggests a practical prescription for NR initial data by starting from a global gauge-consistent metric that is accurate to a given PN order, which could potentially mitigate junk radiation and parameter drift during the early evolution.

%==========================================================================
\begin{acknowledgments}
We thank Saul Teukolsky, Mark Scheel, and Qing Dai for useful discussions.
This work was supported in part by the Sherman Fairchild Foundation,
by NSF Grants
PHY-2309211, PHY-2309231, and OAC-2209656 at Caltech.
Computational resources for this work were supported by Qing Dai.
\end{acknowledgments}

%%%%%%%%%%%%%%%%%%%%%%%%%%%%%%%%%%%%%%%%%%%%%%%%%%%%%%%%%%%%%%%%%%%%%%%%%%%%%%%
%% Uncomment the following line if you want the references set off by
%% a section label instead of the horizontal rule
\def\bibsection{\section*{References}}
\bibliography{References}

\end{document}